\documentclass[12pt]{article}
\usepackage{graphicx}
\usepackage{amsfonts,amssymb}

\usepackage{epstopdf}
\usepackage{cite}
\usepackage{graphics,epsfig}

\usepackage{color}

\def\dalemb#1#2{{\vbox{\hrule height .#2pt
        \hbox{\vrule width.#2pt height#1pt \kern#1pt
                \vrule width.#2pt}
        \hrule height.#2pt}}}

\let\a=\alpha \let\b=\beta \let\g=\gamma \let\d=\delta \let\e=\epsilon
  \let\th=\theta  \let\k=\kappa
\let\l=\lambda \let\m=\mu \let\n=\nu \let\x=\xi  
\let\s=\sigma   \let\f=\phi  
 
\let\w=\omega       \let\D=\Delta \let\Th=\Theta \let\L=\Lambda
\let\X=\Xi  \let\S=\Sigma  \let\Y=\Psi
 
\let\la=\label  
  
\def\nn{\nonumber} \def\bd{\begin{document}} \def\ed{\end{document}}
\def\ds{\documentstyle} \let\fr=\frac \let\bl=\bigl \let\br=\bigr
\let\Br=\Bigr \let\Bl=\Bigl
\let\bm=\bibitem
\let\na=\nabla
\def\tU{{\widetilde U}}
\let\pa=\partial \let\ov=\overline
\def\ie{{\it i.e.\ }}
\newcommand{\be}{\begin{equation}}
\newcommand{\ee}{\end{equation}}
\def\ba{\begin{array}}
\def\ea{\end{array}}
\def\ft#1#2{{\textstyle{{\scriptstyle #1}\over {\scriptstyle #2}}}}
\def\fft#1#2{{#1 \over #2}}
\def\F#1#2{{ F_{#1}^{(#2)} }}
\def\cF#1#2{{ {\cal F}_{#1}^{(#2)} }}

\def\R{{\bf R}}
\def\sst#1{{\scriptscriptstyle #1}}
\def\oneone{\rlap 1\mkern4mu{\rm l}}
\def\e7{E_{7(+7)}}
\def\td{\tilde}
\def\wtd{\widetilde}
\def\im{{\rm i}}
\def\bog{Bogomol'nyi\ }
\newcommand{\ho}[1]{$\, ^{#1}$}
\newcommand{\hoch}[1]{$\, ^{#1}$}
\newcommand{\bea}{\begin{eqnarray}}
\newcommand{\eea}{\end{eqnarray}}
\newcommand{\ra}{\rightarrow}
\newcommand{\lra}{\longrightarrow}
\newcommand{\Lra}{\Leftrightarrow}
\newcommand{\ap}{\alpha^\prime}
\newcommand{\bp}{\tilde \beta^\prime}
\newcommand{\cB}{{\cal B}}
\newcommand{\cO}{{\cal O}}
\newcommand{\vecx}{\vec{x}}
\newcommand{\vecy}{\vec{y}}
\newcommand{\vecp}{\vec{p}}
\newcommand{\vecq}{\vec{q}}
\newcommand{\tr}{{\rm tr} }
\newcommand{\Tr}{{\rm Tr} }
\newcommand{\NP}{Nucl. Phys. }

\newcommand{\cL}{{\cal L}}
\newcommand{\cA}{{\cal A}}

\newcommand{\cD}{{\cal D}}

\def\Ht{\tilde{H}}

\def\sst#1{{\scriptscriptstyle #1}}
\def\0{{\sst{(0)}}}
\def\1{{\sst{(1)}}}
\def\2{{\sst{(2)}}}
\def\3{{\sst{(3)}}}
\def\4{{\sst{(4)}}}
\def\5{{\sst{(5)}}}
\def\6{{\sst{(6)}}}
\def\7{{\sst{(7)}}}
\def\8{{\sst{(8)}}}
\def\9{{\sst{(9)}}}
\def\ve{\varepsilon}
\def\vf{\varphi}
\def\F{\Phi}
\def\wg{\wedge}

\newcommand{\tamphys}{\it 
}

\newcommand{\auth}{AUTHORS}

\def\Lab{\bar{\L}}
\def\Psib{\bar{\Psi}}
\def\Delb{\bar{\D}}
\def\Thb{\bar{\Theta}}
\def\Sigb{\bar{\Sigma}}

\def\sib{\bar{\s}}

\def\mub{\bar{\mu}}
\def\ztb{\bar{\zeta}}
\def\psib{\bar{\psi}}
\def\thb{\bar{\theta}}
\def\chib{\bar{\chi}}

\def\ab{{\bar{a}}}
\def\Ab{\bar{A}}
\def\pb{\bar{p}}
\def\qb{\bar{q}}
\def\cb{\bar{c}}
\def\db{\bar{d}}
\def\vb{\bar{v}}
\def\ub{\bar{u}}
\def\zb{\bar{z}}
\def\Zb{\bar{Z}}

\def\wb{\bar{w}}
\def\lb{\bar{\l}}
\def\Jb{\bar{J}}
\def\Nb{\bar{N}}
\def\pab{\bar{\pa}}
\def\etab{\bar{\eta}}
\def\phib{\bar{\phi}}

\def\e{\epsilon}

\def \bi{\bibitem}
\def \la {\label}

\def \l {\lambda}
\def\foot{\footnote}
\def \tl  {{\tilde \l}}
\def \sql {{\sqrt \l}}
\def \adss {$AdS_5 \times S^5$\ }
\newcommand{\rf}[1]{(\ref{#1})}
\def \ov {\over}

\def\th{\theta}
\def\Th{\Theta}
\def\vth{\vartheta}
\def\btheta{{\bar\theta}}
\def\ttheta{{{\tilde\theta}}}
\def\bttheta{{{\bar\ttheta}}}
\def\vth{\vartheta}

\def\ra{\rightarrow}

\def\au{{\underline{a}}}
\def\bu{{\underline{b}}}
\def\ual{\underline{\a}}
\def\ube{\underline{\b}}
\def\uM{\underline{M}}
\def\uN{\underline{N}}
\def\uP{\underline{P}}
\def\muu{\underline{\mu}}

\def\cc{\circ}
\def\eqv{\equiv}

\def\ni{\noindent}

\def\Ep{E^{{}^{(+)}}}
\def\Em{E^{{}^{(-)}}}

\def\Mp{M^{{}^{(+)}}}
\def\Mm{M^{{}^{(-)}}}

\def \ha{{1\ov 2}}

\def\r{\rho}

\def\Y{{\rm Y}}
\def\X{{\rm X}}
\def\tY{\tilde{\rm Y}}
\def\tX{\tilde{\rm X}}
\def\dY{\dot{\rm Y}}
\def\dX{\dot{\rm X}}

\def \J {\mathcal{J}}
\def \del {\partial}

\def\dF{\dot{F}}
\def\dG{\dot{G}}
\def\adot{\dot{a}}
\def\bdot{\dot{b}}
\def\df{\dot{f}}

\def\dal{{\dot{\alpha}}}
\def\dbe{{\dot{\beta}}}
\def\dga{{\dot{\gamma}}}

\def \E {{\cal E}}
\def \cS {{\cal S}}
\def \J {{\cal J}}
\def\N{{\cal N}}
\def\F{{\cal F}}
\def\M{{\cal M}}

\def\ms{\mathcal{S}}
\def\mj{\mathcal{J}}
\def\soj{\fr{\ms}{\mj}}
\def \R {{\bf R}}
\def \om {\omega}
\def \bE {\bar E}
\def \x {{\cal X}}

\def \bi{\bibitem}
\def \la {\label}

\def \l {\lambda}
\def\foot{\footnote}
\def \tl  {{\tilde \l}}
\def \sql {{\sqrt \l}}
\def \adss {$AdS_5 \times S^5$\ }
\def \ov {\over}

\def \varpi {{\rm w}}

\def\e{\epsilon}

\def\At{\tilde{A}}
\def\Bt{\tilde{B}}
\def\Ct{\tilde{C}}
\def\Dt{\tilde{D}}
\def\Et{\tilde{E}}
\def\Ft{\tilde{F}}
\def\Gt{\tilde{G}}
\def\Mt{\tilde{M}}
\def\at{{\tilde{a}}}
\def\bt{{\tilde{b}}}
\def\ct{\tilde{c}}
\def\dt{\tilde{d}}
\def\et{\tilde{e}}
\def\ft{\tilde{f}}
\def\gt{\tilde{g}}
\def\chit{\tilde{\chi}}
\def\psibt{\tilde{\psib}}
\def\mut{\tilde{\mu}}
\def\mubt{\tilde{\mub}}
\def\wt{\tilde{w}}
\def\wbt{\tilde{\wb}}

\def\ola{\overleftarrow}
\def\ora{\overrightarrow}
\def\alt{\tilde{\a}}

\def\dh{\hat{d}}
\def\bh{\hat{b}}
\def\deh{\hat{\d}}
\def\mh{\hat{m}}
\def\nh{\hat{n}}

\def\ah{\hat{a}}
\def\eh{\hat{e}}
\def\Eh{\hat{E}}
\def\eph{\hat{\e}}
\def\ph{\hat{p}}
\def\Ah{\hat{A}}
\def\alh{\hat{\a}}
\def\beh{\hat{\b}}
\def\gah{\hat{\g}}
\def\muh{{\hat{\m}}}
\def\nuh{{\hat{\n}}}
\def\thh{\hat{\th}}
\def\dh{\hat{d}}
\def\ih{\hat{i}}
\def\jh{\hat{j}}
\def\deh{\hat{\d}}
\def\uh{\hat{u}}
\def\vh{\hat{v}}
\def\wh{\hat{w}}
\def\lah{\hat{\l}}
\def\Ch{\hat{C}}
\def\Omh{\hat{\Omega}}
\def\rh{\hat{r}}
\def\sh{\hat{s}}
\def\that{\hat{t}}
\def\lah{\hat{\l}}

\def\clb{\color{blue}}
\def\clr{\color{red}}
\def\clg{\color{green}}

\def\dgg{\dagger}

\def\ps{\rlap{\, /}\;\,p }
\def\ks{\rlap{\, /}\;\,k }
\def\Ds{\rlap{\, /}\;\,D }
\def\Hs{\rlap{\, /}\;H }

\def\gym{g_{YM}}

\def\jbf{\mathbf{j}}
\def\kbf{\mathbf{k}}
\def\ybf{\mathbf{y}}
\def\xbf{\mathbf{x}}
\def\Lbf{\mathbf{L}}
\def\rbf{\mathbf{r}}

\def\bpa{\bar{\pa}}

\newcommand{\bes}{\begin{subequations}}
\newcommand{\ees}{\end{subequations}}

\newcommand{\eda}{\bar{\eta}}
\newcommand{\teta}{\tilde{\eta}}
\newcommand{\tta}{\tilde{\eta}}
\newcommand{\sP}{/\!\!\!\partial}
\newcommand{\sF}{\: /\!\!\!\! F}

\newcommand{\dd}{{d}}

\newcommand{\dsfrac}{\displaystyle\frac}

\begin{document}
\overfullrule=0pt
\parskip=2pt
\parindent=12pt
\headheight=0in \headsep=0in \topmargin=0in
\oddsidemargin=0in

\vspace{ -3cm}
\thispagestyle{empty}



 \vspace{0.1cm}

\setcounter{equation}{0}
\setcounter{footnote}{0}
\setcounter{section}{0}



\begin{center}

{\Large\bf On the pattern of black hole information release
  }

\vskip 0.8cm



\vspace{0.5cm}
I. Y. Park
\\

{\it Center for Quantum Spacetime, Sogang University\\
Shinsu-dong 1, Mapo-gu, 121-742 South Korea \\
}

and

\vspace{0.1cm}
{\it Department of Natural and physical Sciences,
Philander Smith College 
                               \\
Little Rock, AR 72223, USA \\
inyongpark05@gmail.com
}

\end{center}

 \vspace{0.1cm}

 \begin{abstract}

We propose a step towards a resolution to black hole information paradox by
analyzing scattering amplitudes of a complex scalar field around a Schwarzschild black hole. The
scattering cross section reveals much information on the incoming state but exhibits
flux loss at the same time. The flux loss should be temporary, and indicate mass growth of the black hole. The black hole should Hawking-radiate subsequently, thereby, compensating for the flux loss. {By examining the purity issue}, we comment on
the possibility that information bleaching may be the key to the paradox.

\end{abstract}
\newpage

\setcounter{equation}{0}
\setcounter{footnote}{0}
\setcounter{section}{0}

\section{Introduction}

Ever since the discovery of Hawking radiation \cite{Hawking:1974sw}, Black Hole (BH) Information paradox \cite{Hawking:1976ra} has remained a puzzle although there has been some progress. (See, e.g., \cite{Davies:1978zz,Birrell,Wald,Wipf:1998ss,Carroll,Susskindbook,Jacobson:2003vx,Muk,Parker} for reviews on BH physics.)
There have been various approaches to a resolution of the paradox \cite{Susskind:1993if,Stephens:1993an,'tHooft:1996tq,Parikh:1999mf,Burgess:2003jk,Amati:2007ak,Englert:2010cg,
Mathur:2009hf,Skenderis:2008qn,Zhang:2009jn,Park:2009mi,Czech:2011wy,Avery:2011nb,Giddings:2012dh}.
A new proposal \cite{Almheiri:2012rt} (see \cite{Braunstein:2009my} for an earlier related discussion) appeared recently and a number of works \cite{Bousso:2012as,Mathur:2012jk,Banks:2012nn,Hwang:2012nn,Susskind:2012uw,Avery:2012tf,Larjo:2012jt,Page:2012zc,
Papadodimas:2012aq,Nomura:2012sw,Susskind:2012rm,Brustein:2012jn,Rama:2012fm} have followed.

{ In this work, we approach the information problem from a very different angle by
{touching upon} the issue whether an infalling object carries its information into the
black hole inside. (Another crucial issue is whether the information stay inside while the black
hole grows, and we will come back to this later.) It seems that absence of information bleaching at the horizon has been a widely accepted assumption \cite{Preskill:1992tc}. We do not believe that this assumption has ever been proven outside the realm of the semiclassical description, and its invalidity may well lead to a solution of the information puzzle.

If a certain information bleaching is present, its impact on the information issue will be dramatic.\footnote{{The author has become aware of the observation made in \cite{Susskind:1993if} only very recently (long after the first version of the present paper appeared in the web). Below eq(1.1) they stated, ``... so we arrive at the conclusion that the state inside the black hole, $|\Phi(\S_{bh})>$, must be independent of the initial state. In other words, all distinctions between initial states of infalling matter must be obliterated before the state crosses the global event horizon." We anticipate that the obliteration in the semi-classical description should translate into bleaching once the quantum field theory interactions are taken into account.}}
 Suppose neutral massless scalar particles (we consider a particle instead of a chair {\em for simplicity}) approaching a black hole from far away. The information that they carry is velocity and the amount, i.e., flux.
Let us also assume that the particle dynamics is described by an {\em interacting} quantum field theory or quantum gravity, but not a free theory. The question that has motivated the current work was: would those pieces of information enter the black hole? If so, would they remain inside, during the thermalization process? The semiclassical description should not be adequate for tackling these questions. This is because a particle with a quantum field theory interaction would produce jets as it approaches the black hole, and the jet will have a characteristic pattern just like in real QCD experiments. The characteristic pattern will revel information about the particle. Could the jet production be such that, for example, the information on the velocity gets bleached and only the information on the amount enters the black hole?

We believe that this is a genuine possibility, and, if true, it would signify that only the minimal information (i.e., the infrmation consistent with no-hair theorem), such as information on the amount/flux, enters the black hole. (Or if it enters, it might escape while black hole grows. (See later discussions.))
 If present, a bleaching mechanism would have a drastic effect on our strategy for solving BH information: we would find out what information actually entered the BH before worrying about loss of that information.
 {The jet production would affect the purity consideration. The whole system is now consider as the Hawking radiation, "non-Hawking radiation", jets and/or the modes behind the horizon. The entire system might never become thermal at any point of time but be always pure. In other words, it might be only the original Hawking radiation that is thermal. The fact that the Hawking radiation is thermal does not mean that the entire system is thermal because Hawking radiation is part of the system. Much of the information would be embedded in the jets which approach the event horizon. (The approaching jets contain information, and might be related to the concept of stretched horizon.)} We believe that scattering around a black hole should provide an essential tool\footnote{{ Computation of scattering amplitudes was carried out, e.g., in \cite{Verlinde:1993sg} in the context of the CGHS model.
}}, and take a step in this work.

}

As wellknown, the vacuum for an infalling observer - known as the Kruskal vacuum - appears as a thermal state to a stationary observer. By the same token, one can study the way in which the scattering of one-particle states built on
the Kruskal vacuum will appear to a stationary observer.
Motivated in part by simplicity, we employ a free complex scalar\footnote{
{  Although a free quantum field theory should be largely inadequate, there are tasks such as the ones in this work that can be
carried out. (Also, an interacting quantum field theory setup was unavailable at the time of writing this article; it has been recently obtained in \cite{Park:2013iqa},\cite{Park:2013vpa} and \cite{appear}.)}
} in two dimensions to focus on the essential aspects of the analysis.

The theory is considered in the ``two-dimensional
Schwarzschild black hole" background. We study several scattering amplitudes, and interpret the results in light of black hole information paradox. As we will show in this work, the amplitudes exhibit flux loss.\footnote{Flux loss was discussed in \cite{Park:2009mi} in the string theory vertex operator formulation.}
The potential resolution of the information paradox that we propose is such that
 there should be ``radiation" that is emitted before the black hole becomes completely thermalized, and this part of the radiation should be non-thermal, i.e., information-containing.
 The flux loss should imply BH's gain of mass from the incoming wave.
 After the black hole is thermalized, it will Hawking-radiate thermally.
This implies, in particular, that the information loss should only be an apparent phenomenon due to different timing of two different ``radiations", thermal and non-thermal.

There exists a subtle tension between the quantum mechanical description and the quantum field theoretical
description of the physics at hand.\footnote{{Quantum field theory is based on quantum principle, of course. Distinguishing quantum field theory from quantum mechanics is to stress the multi-particle scattering capability of the former. The multi-particle nature of quantum field theory was essential for Hawking to establish Hawking temperature in his original work.}} On one hand, one must deal with scattering by the potential produced
by a black hole, and the quantum mechanical description seems adequate to that end. On the other hand, the finite temperature inducing effect of a BH was established in the quantum field theory framework. We note below that the essential part of the quantum mechanical analysis can be integrated into the quantum field theory setup. The connection is made by the role that the currents play in the quantum mechanics and quantum field theory.

We point out that the cross section in a curved spacetime should be viewed as a
position-dependent quantity in general. The reason is that each point in the space has its own
phase space that may be different from that of the others. The notion of
 position-dependent cross section might not be useful in a spacetime of generic geometry; however, the Schwarzschild geometry has a sufficient amount of symmetry for the notion to be well-defined and useful.

\vspace{.2in}

The organization of the remainder is as follows:
In section 2, we first review the relationship between flux and differential cross section
in the flat space non-relativistic quantum mechanics (QM) setup. The differential cross section
is related to the current in the standard manner. Then we review the quantum field theory (QFT) discussion that led to Hawking radiation. This provides
a useful guide in section 3 in which we employ a QFT setup to tackle scattering of one-particle states around a
Schwarzschild black hole. Although the discussion in section 3 concerns two-dimensional spacetime, the essential results can be extended to 4D as we will discuss after the 2D case.
In retrospect of our result, we speculate on the potential relevance of the information bleaching, and comment on what it would mean for a chair thrown into a BH. We conclude with a summary and future directions.

\section{Review of QM- and QFT- setups}

In a "global" view, we propose treating the presence of
a black hole as a target of scattering particles.
We start with a review of quantum mechanics (QM) analysis of scattering around a target with spherical symmetry in the
next subsection. This will help us get oriented in the following sections in which the quantum field theory (QFT) analysis
of scattering around a BH is carried out.

\subsection{Scattering in non-relativistic QM}

One way to introduce a differential cross section in QM is to start with flux.
 We review this procedure by closely following \cite{Gasi}.
Let us consider a quantum mechanical system of a particle of mass $m$ with spherical symmetry. The asymptotic form of the
wave-function can be written as
\bea
\psi(\mathbf{r})\simeq e^{i\kbf \cdot\rbf}+f(\th)\fr{e^{ikr}}{r}
   \la{awf}
\eea
The flux is defined by
\bea
\jbf=\fr{1}{2im}(\psi^*\nabla \psi-\psi\nabla \psi^*) \la{nrf}
\eea
Substituting \rf{awf} into \rf{nrf}, one finds
\bea
\jbf&=&\fr{ \kbf}{m}+\fr{ k}{m}\,\hat{i}_r |f(\th)|^2 \fr1{r^2}\nn\\
  && +\fr{ \kbf}{2m}\;\fr1{r}\big[f^*(\th)e^{-ikr(1-\cos\th)}+f(\th)e^{ikr(1-\cos\th)} \Big]\nn\\
  &&+\fr{ k}{2m}\;\fr{\hat{i}_r}{r}\big[f^*(\th)e^{-ikr(1-\cos\th)}+f(\th)e^{ikr(1-\cos\th)} \Big]\nn\\
  &&-\fr{ 1}{2im}\;\fr{\hat{i}_r}{r^2}\big[f(\th)e^{ikr(1-\cos\th)}-f^*(\th)e^{-ikr(1-\cos\th)} \Big]\nn\\
 &&+\fr{ 1}{2im}\;\fr{\hat{i}_\th}{r^2}\big[\fr{\pa f(\th)}{\pa\th}e^{ikr(1-\cos\th)}-\fr{\pa f^*(\th)}{\pa\th}e^{-ikr(1-\cos\th)} \Big]  \la{qmj}
\eea
where $\hat{i}_r$ denotes a unit vector along $\rbf$, we have used $\kbf\cdot \rbf=kr\cos \th$ in the exponents, and omitted $\fr1{r^3}$ terms. Let us label the terms in the first line as
\bea
\jbf_{1}\equiv \fr{ \kbf}{m}\quad,\quad
\jbf_{2}\equiv \fr{ k}{m}\,\hat{i}_r |f(\th)|^2 \fr1{r^2}
\eea
The differential cross section $d\s$ is defined by
\bea
d\s= \fr{\jbf_2}{|\jbf_{1}|}\cdot \hat{i}_r dA
\eea
where $dA$ is the surface element that is perpendicular to the radial direction.
Only the $\jbf_2$ term genuinely contributes to the asymptotic cross section. The $\jbf_{1}$ term is disregarded
by not counting the forward scattering.
Since the area element in 3D space is $dA\sim r^2$, the flux is conserved and
the cross section has a non-vanishing value. We will see in the next section that this is not the case for scattering around a Schwarzschild BH.

\subsection{Review of black hole thermal radiation}

To set the stage for QFT amplitude computations in section \ref{main}, let us review basics of
a free scalar field theory in a 2D ``BH" background.
Consider a complex scalar action
\bea
S= -\int d^2 x\sqrt{-g} \;\pa_\m\f^* \pa^\m \f  \la{2Daction}
\eea
The case of a real scalar field was discussed, e.g., \cite{Birrell,Carroll} whose conventions we follow with minor changes. (See, e.g., \cite{Dimock:1987hi} for a treatment with mathematical rigor.) We obtain complex versions of the corresponding equations in \cite{Birrell} in this subsection.

There are three main sets of coordinates that we use:
\bea
&& (\ub,\vb):\;\; \mbox{Kruskal-Szekeres coordinates (or Kruskal coordinates for short)}  \nn\\
&& (u,v):\;\; \mbox{tortoise-null coordinates}\;\; \mbox{(in some places tortoise coordinates $(t,\rh)$ are used)} \nn\\
&&  (t,r)  :\;\; \mbox{Schwarzschild coordinates}
\eea
The relations between tortoise-null coordinates $(u,v)$ and Kruskal coordinates are
\bea
\ub &=& -4M e^{-\fr{u}{4M}}   \nn\\
\vb &=&  4M e^{\fr{v}{4M}}
\eea
where
\bea
u\equiv t-\rh,\quad v\equiv t+\rh,\quad \rh\equiv r+2M\ln |\fr{r}{2M}-1|
\eea
 The metric for the 2D Schwarzschild BH is given by
\bea
ds^2= -\fr{2M}{r}e^{-\fr{r}{2M}} d\ub d\vb
 \la{Kmetric}
\eea
in the Kruskal coordinates, by
\bea
ds^2= -\fr{2M}{r}e^{-\fr{r}{2M}} e^{\fr{v-u}{4M}}dudv
\eea
in the tortoise-null coordinates $(u,v)$.
The mode expansion of $\f$ in the tortoise-null coordinates is given by
\bea
\f=\sum_{k=-\infty}^{\infty}\Big[b_{+}^L(k) u_{k}^L+b_{-}^L(k)^{\dagger} u_{k}^{L*}
    + b_{+}^R(k) u_{k}^R+b_{-}^R(k)^{\dagger} u_{k}^{R*}\Big]   \la{rfcorrect}
\eea
The functions $u_k^{L,R}$ are defined as
\bea
u_k^R =\left\{ \begin{array}{c}
        \fr1{\sqrt{4\pi \w}}e^{-i\w t+ik \rh},\quad \mbox{in}\;\; R \\
        \;\;0, \hspace{1in} \mbox{in}\;\; L
         \end{array}
      \right. \la{ukr}
\eea
\bea
u_k^L =\left\{ \begin{array}{c}
        \fr1{\sqrt{4\pi \w}}e^{i\w t+ik \rh},\quad \mbox{in}\;\; L \\
        \;\;0, \hspace{1in} \mbox{in}\;\; R
         \end{array}
      \right.    \la{ukl}
\eea
The reason for introducing two sets of $u_k$'s is that the positive frequency functions - with which the annihilation operator is associated - take different forms in the regions L and R.
The mode expansion of $\f$ in the Kruskal coordinates can be written as
\bea
\f=\sum_k \Big[c_+^L(k) v_k^L+c_-^L(k)^\dagger v_k^{L*}
     + c_{+}^R(k) v_{k}^R+c_{-}^R(k)^{\dagger} v_{k}^{R*}\Big]
\eea
where $v_k^L,v_k^R$ are related to $u_{k}^L,u_{k}^R$ by:
\bea
v_k^L &=&  \fr1{\sqrt{2\sinh \fr{\pi \w}{a}}}\Big(
  e^{\fr{\pi \w}{2a}} u_{k}^L+e^{-\fr{\pi \w}{2a}} u_{-k}^{R*}
  \Big)\nn\\
v_k^R &=&\fr1{\sqrt{2\sinh \fr{\pi \w}{a}}}\Big(
  e^{\fr{\pi \w}{2a}} u_{k}^R+e^{-\fr{\pi \w}{2a}} u_{-k}^{L*}
  \Big)
\eea
The $b$- and $c$- oscillators are related by a similarity transformation \cite{Unruh:1976db,Israel:1976ur}:
\bea
b_+^L(k)=e^{iB_1}c_+^L(k) e^{-iB_1}, \quad b_+^R(k)=e^{iB_1}c_+^R(k) e^{-iB_1}
\la{bcrel}
\eea
where
\bea
B_1 &\equiv & \sum_k i\f_\w \Big[b_+^{L}(-k)^\dagger b_-^{R}(k)^\dagger-b_+^{L}(-k)b_-^{R}(k)\Big]
  \nn\\
B_2 &\equiv & \sum_k i\f_\w \Big[b_-^{L}(-k)^\dagger b_+^{R}(k)^\dagger-b_-^{L}(-k)b_+^{R}(k)\Big]
\la{bpbm}
\eea
with
\bea
\tanh \f_\w\equiv e^{-\pi \w /a} \la{fo}
\eea
The $b_-,c_-$ oscillators satisfy similar relations.
These relations imply
\bea
&&b_+^L(k)=\fr{1}{\sqrt{2\sinh (\pi\w/a)}}\Big[
                         e^{\pi \w/2a}c_+^L(k)  +e^{-\pi \w/2a}c_-^{R}(-k)^{\dagger}
                           \Big]  \nn\\
&&b_+^R(k)=\fr{1}{\sqrt{2\sinh (\pi\w/a)}}\Big[
                         e^{\pi \w/2a}c_+^R(k)  +e^{-\pi \w/2a}c_-^{L}(-k)^\dagger
                           \Big]
                           \la{boscill}
\eea
Inverting \rf{boscill}, one finds
\bea
c_+^L(k) &=& \fr{1}{\sqrt{2\sinh (\pi\w/a)}}\Big[
                         e^{\pi \w/2a}b_+^L(k)  -e^{-\pi \w/2a}b_-^{R}(-k)^\dagger
                           \Big]  \nn\\
c_-^{R}(k) &=& \fr{1}{\sqrt{2\sinh (\pi\w/a)}}\Big[
                         -e^{-\pi \w/2a}b_+^L(-k)^\dagger  +e^{\pi \w/2a}b_-^{R}(k)
                           \Big]
\eea
The Kruskal vacuum $|0,K>$ and the tortoise-null vacuum $|0,S>$ are connected by the following equation that follows from \rf{bcrel},
\bea
|0,K>=e^{-iB_2}e^{-iB_1}|0,S>
\eea
 The $|0,K>$ vacuum can be expressed as\footnote{This is the complex version of the corresponding equations in the real scalar case \cite{Unruh:1976db}\cite{Israel:1976ur}:
\bea
|0,K>
&=& e^{\sum_k [-\ln \cosh \f_{\w}+i\tanh \f_{\w}\, {b^{L}(k)^\dagger} {b^{R}(k)^\dagger}]}\;|0,S>\nn\\
&=& \Big(\prod_{k'} \fr{1}{\cosh \f_{\w_{k'}}}\Big)
 \Big[  \prod_{\w_k}\sum_{n_k=0}^\infty \; (i)^{n_k}\,e^{-\fr{n_k\; \pi \,\w}{\k} }
     |n_{kL}>|n_{kR}> \Big]
\eea
 These are effective equalities: the full equalities contain an
 additional factor
\bea
e^{-\fr{i\pi }{4}[b^{L}(-k)^\dagger b^{L}(k)+b^{R}(-k)^\dagger b^{R}(k)] }
\eea
that does not affect the expectation values of the number operators.
 }
\bea
|0,K>
&=& e^{\sum_{k'} [-\ln \cosh \f_{\w'}+i\tanh \f_{\w'}\, {b_-^{L}(k')^\dagger} {b_+^{R}(k')^\dagger}]}\;
    e^{\sum_k [-\ln \cosh \f_\w+i\tanh \f_\w\, {b_+^{L}(k)^\dagger} {b_-^{R}(k)^\dagger}]}|0,S>\nn\\
&=& \Big(\prod_{k''} \fr{1}{\cosh \f_{\w_{k''}}}\Big)^2
 \Big[ \prod_{\w_{k'}}\sum_{n_{k'}=0}^\infty (i)^{n_{k'}}\,e^{-\fr{n_{k'}\; \pi \,\w}{\k} }
      |n_{k'L}^{- }>|n_{k'R}^{+}>\Big]\nn\\
 &&\hspace{1.2in}
 \Big[  \prod_{\w_k}\sum_{m_k=0}^\infty (i)^{m_k}\,e^{-\fr{m_k\; \pi \,\w}{\k} }
     |m_{kL}^{+ }>|m_{kR}^{- }> \Big]
\eea
where
\bea
\k=\fr1{M}
\eea
It is somewhat puzzling that the state $|0,K>$ has particles and antiparticles populated in the same manner.
If we were using a realistic theory with QFT interactions, this would mean that $|0,K>$ is a highly unstable state from the standpoint of a Schwarzschild observer.
Let us define the $R$-region number operators in the usual manner,
\bea
N_+^R(k)= b_+^{R}(k)^\dagger \;b_+^{R}(k)  \quad,\quad N_-^R(k)= b_-^{R}(k)^\dagger \;b_-^{R}(k)
\eea
and compute $<0,K|N_+^R(k)|0,K>$:
\bea
<0,K|N_+^R|0,K>&=& \Big(\prod_{q} \fr{1}{\cosh \f_{\w_{q}}}\Big)^2
      \Big[   <n_{p'L}^{- }|<n_{p'R}^{+ }| \prod_{\w_{p'}}\sum_{n_{p'}=0}^\infty e^{-\fr{n_{p'}\; \pi \,\w_{p'}}{\k} }\Big]
\nn\\
 && \hspace{1in}   N_+^R(k)  \Big[  \prod_{\w_p}\sum_{n_p=0}^\infty e^{-\fr{n_p\; \pi \,\w_{p}}{\k} }
     |n_{pL}^{- }>|n_{pR}^{+ }> \Big]\nn\\
&=& (1-e^{-2\pi \w_k/\k}) \sum_{n=0}^\infty ne^{-2\pi n\w_k/\k}    \nn\\
&=&\fr{1}{e^{{2\pi M \w_{_k}}}-1}   \la{na}
\eea

\section{Scattering around a black hole: a QFT account \la{main}}

In this section, we ponder how to integrate the QM ingredients\footnote{See, e.g., \cite{Sanchez:1977si,matz} for works that employed the QM setup.} of the
previous section into the quantum field theory setup. We first focus on the free complex scalar theory
in two dimensions, and briefly discuss how things should be in higher dimensions.

In the conventional approaches, a black hole is treated as a classical background while the matter field (such as a scalar) is quantized.
One includes various matter fields for an extended analysis along this line. Inclusion of various fields, however, does not automatically establish that the real black
hole emits quanta of those fields. For that, one must make sure that the fields are the intrinsic degrees of freedom of the black hole. 

We have proposed a new paradigm for black hole physics in our recent work \cite{Hatefi:2012bp} (that has been further developed in \cite{Park:2013iqa},\cite{Park:2013vpa} and \cite{appear}):
 the DBI action that results from carrying out ADM decomposition followed by Hamilton-Jacobi procedure
 \cite{Sato:2002kv} in the
supergravity action that is obtained by expanding around a black hole solution. One of the nice features of this approach is that the entire procedure can be carried out in a pure Einstein gravity (as opposed to, e.g., a higher dimensional supergravity).
As mentioned above, a scalar field
is considered in the black hole {\em background} in the conventional approach, and this makes the scalar field extrinsic to the BH unless one fully quantizes the coupled system. Using a field that is {\em intrinsic} degrees of the black hole would be far more ideal, given the lack of gravity quantization.

In principle, we should start with the ADM approach, and use the precise form of the theory on
 a hypersurface \cite{Hatefi:2012bp}. Instead we consider a 2D complex scalar theory in this work, and therefore the vintage point of the ADM reduction is limited to the conceptual aspect for now.\footnote{ The proper setup has been developed now in \cite{Park:2013iqa},\cite{Park:2013vpa} and \cite{appear}.}

We focus on an eternal (i.e., preexisting) BH rather than a gravitational collapse, and consider scattering particles against a black hole target.
We compute several quantities below to analyze flux loss and information release.
One such quantity is an expectation value of
flux or the $U(1)$ symmetry current $j_\m$.
The flux loss should indicate an increase in BH mass. The scattering amplitudes exhibit information on the incoming particles.
The subsequent decay of the BH would be thermal but would not lead to loss of information.
Only part of the particle flux (but nothing else) is temporarily lost.

As in the QM case, we examine the cross section to determine flux loss.
As stated before, the cross section in a curved spacetime should in general be viewed as a
position-dependent quantity.
The main motivation for considering the current is
to define ``the finite distance cross section" as against the asymptotic
cross section. We introduce this cross section in the next subsection.
 For the flux-preserving cases, the finite-distance cross section and the usual asymptotic cross section are the same. For the flux-losing cases, however, the former is a decreasing function of distance with the minimum value taken by the asymptotic cross section.

\subsection{computation of amplitudes \la{ampcomp}}

We have reviewed how the current is related to the differential cross section in the QM formulation
in section 2.
The quantum field theoretic cross section is naturally defined when there are two incoming particles \cite{Weinberg}\cite{Sterman}. The QFT amplitudes that we consider
 describe amplitudes of a {\em one}-particle state scattered around the potential generated by the BH. The differential cross section can still be defined in analogy with the QM
formulation.
The U(1) symmetry current associated with the action \rf{2Daction} is given by
\bea
j_\m(x)=i\,[\pa_\m \f^*(x)\,\f(x)-\f^*(x)\,\pa_\m \f(x)] \la{crdcu}
\eea
where $x$ represents the 2D tortoise coordinates,
\bea
x^\m \equiv (t,\rh)
\eea
Upon substituting the mode expansion, the current \rf{crdcu} takes, for $x\in R$ region,
\bea
j_\m(x) &=& \sum_{q,q'}j_{\m, qq'}(x)
\eea
where
\bea
j_{\m, qq'}(x)&\equiv & 2i\Big[
           b_{+}^R(q)^\dagger  b_{+}^R(q') u_{q'}^{R}\pa_\m u_q^{R*}
          +  b_{-}^R(q) b_{-}^R(q')^\dagger u_{q'}^{R*}\pa_\m u_q^{R}\nn\\
   &&\hspace{.4in}+b_{+}^R(q)^\dagger  b_{-}^R(q')^\dagger u_{q'}^{R*}\pa_\m u_q^{R*}
          +  b_{-}^R(q) b_{+}^R(q') u_{q'}^{R}\pa_\m u_q^{R}
    \Big]  \la{jinb}
\eea
Let us compute $<0,K|j_\m(x)|0,K>$. The current is only sensitive to the net number of the particles or antiparticles.
Since the particles and anti-particles are populated in the same manners in $|0,K>$, it is expected that
 the result of the computation will vanish. (However, one can separate out the particle contribution
 from that of the anti-particles as we will discuss below.) The third term and the fourth term in \rf{jinb}
are analogous to the QM discussion terms in the second to fourth lines in \rf{qmj}. The terms
 would vanish due to Riemann-Lesbegue lemma when integrated over the angles in higher dimensions and therefore will not be considered.

The first term can easily be computed as a minor variation of \rf{na}:
\bea
 &&<0,K|b_+^R(q)^\dagger b_+^{R}(q')|0,K>   \nn\\
 &&= \Big(\prod_{q''} \fr{1}{\cosh \f_{\w_{q''}}}\Big)^2
      \Big[   <n_{p'L}^{- }|<n_{p'R}^{+ }| \prod_{\w_{p'}}\sum_{n_{p'}=0}^\infty e^{-\fr{n_{p'}\; \pi \,\w_{p'}}{\k} }\Big]   \nn\\
 && \hspace{1in}  b_+^R(q)^\dagger b_+^{R}(q')
  \Big[  \prod_{\w_p}\sum_{n_p=0}^\infty e^{-\fr{n_p\; \pi \,\w_{p}}{\k} }
     |n_{pL}^{- }>|n_{pR}^{+ }> \Big]\nn\\
  &&= \Big(\prod_{q''} \fr{1}{\cosh \f_{\w_{q''}}}\Big)^2
      \Big[   <n_{p'L}^{- }|<n_{p'R}^{+ }| \prod_{\w_{p'}}\sum_{n_{p'}=0}^\infty e^{-\fr{n_{p'}\; \pi \,\w_{p'}}{\k} }\Big]   \nn\\
 && \hspace{1in}  \d^\2({q}-{q}')N_+^R(q)
  \Big[  \prod_{\w_p}\sum_{n_p=0}^\infty e^{-\fr{n_p\; \pi \,\w_{p}}{\k} }
     |n_{pL}^{- }>|n_{pR}^{+ }> \Big]\nn\\
 &&=\d^\2({q}-{q}')\fr{1}{e^{{2\pi M}\w_q}-1}
  \la{bbd}
\eea
Let us rewrite the second term in \rf{jinb} as
\bea
&& <0,K|b_{-}^R(q) b_-^{R}(q')^\dagger|0,K>\nn\\
 =&& <0,K|\d^\2({q}-{q}')|0,K>
                           +<0,K|b_-^{R}(q')^\dagger b_-^R(q) |0,K>
   \la{bbd2}
\eea
The delta-function (or Kronecker delta) term may be dropped by normal ordering in analogy with the flat case:
\bea
<0,K|:b_{-}^R(q) b_-^{R}(q')^\dagger:|0,K>
=\d^\2({q}-{q}')\fr{1}{e^{{2\pi M}\w_q}-1}  \la{bmbmd}
\eea
This result, combined with \rf{bbd}, establishes $<0,K|j_\m(x)|0,K>=0$ as anticipated.
It is useful for our purpose, however, to separate out the particle contribution from the anti-particle contribution.
One may also pick out a given mode $k$; this amounts to introducing a field $\f_+(k)$,
\bea
\f_{+,k}(x)= b_{+}^L(k) u_{k}^L(x)
    + b_{+}^R(k) u_{k}^R(x)   \la{F}
\eea
and its current for $x\in R$,
\bea
j_{\m,k}^+(x)\equiv i\,[\pa_\m \f_{+,k}^*\,\f_{+,k}-\f_{+,k}^*\,\pa_\m \f_{+,k}]
\eea
The amplitude for this current is given by
\bea
<0,K|j_{\m,k}^+(x)|0,K> &=& i\,<0,K|[\pa_\m \f_{+,k}^*\,\f_{+,k}-\f_{+,k}^*\,\pa_\m \f_{+,k}]|0,K> \nn\\
          &=& 2i\, u_{k}^{R}(x)\pa_\m u_k^{R*}(x) <0,K|
           b_{+}^R(k)^\dagger  b_{+}^R(k) |0,K> \nn\\
          &=&   2i\, u_{k}^{R}(x)\pa_\m u_k^{R*}(x) \fr{1}{e^{{2\pi M}\w_k}-1}
          \la{jk}
\eea
and this represents the particle contribution.
(We will come back to a related point towards the end of this subsection.)
The factor $ \fr{1}{e^{{2\pi M}\w_k}-1}$ represents the effect of the finite temperature background.
 Now one may examine the total flux that passes through a surface of a given radius by multiplying
 the surface element and integrating over the angle. It will be just the metric factor in the case of 2D without the surface integration since the "surface element" is a point.

The expression \rf{bbd} represents the amplitude measured in the tortoise-null coordinates.
One can convert \rf{jk} into the corresponding expression
in the Schwarzschild coordinates, $x'^\m\equiv (t,r)$
\bea
<0,K|j_{r,k}^{+'}(x')|0,K> &=& \fr{\pa x^\n}{\pa x'^r} <0,K|j_{\n,k}^+(x)|0,K>  \nn\\
                            &=& 2i\, u_{k}^{R}(x)\pa_{r^*} u_k^{R*}(x) \fr{1}{e^{{2\pi M}\w_k}-1}
                                \left[1+\fr{2M}{r}\right]
                               \la{jt}
\eea
where we have used $\fr{\pa \rh}{\pa r}\simeq 1+\fr{2M}{r}$ in the second equality.
Multiplying the zero-dimensional surface element $dA_0\simeq 1+\fr{2M}{r}$ in the Schwarzschild coordinates, one gets
\bea
  <0,K|\;j_{r,k}^{+'}(x')|0,K> dA_0 &=& 2k\,  \fr{1}{e^{{2\pi M}\w_k}-1}
                                \left[1+\fr{4M}{r}\right]  \la{jp}
\eea
  The leading piece is analogous to the $\mathbf{j}_1$ term of the QM discussion, and should be associated with the incoming flux.
  The subleading term is associated with the outgoing flux, and thus it should play a role analogous to the $\jbf_{2}$ current in the QM discussion in sec 2.\footnote{In the QM case, one can simply avoid the forward direction when measuring the scattered wave. In the higher dimensional analogue of the present case, one can place the detector at locations farther away than the source of the spherical incoming wave. One can adjust the width of the incoming beam of the plane-wave
  in the QM case. By the same token, it should be possible to adjust the incoming spherical wave to exist within a certain radius in the higher-dimensional analogue of the current case.}
A two-dimensional observer will be able to deduce the incoming wave by measuring
the scattered wave. The only thing that is lost temporarily is the amount of the flux. The observer
will also have to measure Hawking radiation in order to correctly recover the information on the incoming flux.

An interesting aspect of the flux is revealed by checking a potential anomaly in the continuity equation:
\bea
\nabla_\m <0,K|j^\m(b)|0,K>=\fr1{\sqrt{g}}\pa_\m (\sqrt{g}\,<0,K|j^\m(b)|0,K>)  \la{coneq}
\eea
This vanishes in the tortoise-null coordinates, whereas in the Schwarzschild coordinates it leads to an expression that contains $O\Big(\fr{1}{r^2}\Big)$. This may have a deeper meaning with regards to how the Kruskal coordinates
deal with a singularity.

Now let us consider the following scattering amplitude in analogy with the flat case,
\bea
&& <0,K|b_+^R(k') b_+^{R}(k)^\dagger|0,K> =\d^\2({k}-{k}')\fr{e^{{2\pi M}\w_k}}{e^{{2\pi M}\w_k}-1}
   \la{bbd3}
\eea
The advantage of considering this amplitude, compared with \rf{jp}, is that the presence of the $\d^\2({k}-{k}')$ factor clearly shows {\em the unitary nature of the scattering}. {\em This is because the amplitude exhibits  dependence on the state of the incoming particle}, the momentum.\footnote{Informally speaking, ``what comes" out depends on ``what enters".}
The disadvantage is that the transformation rule from the tortoise-null coordinates to the Schwarzschild coordinates is obscured unlike the case of the current discussed in \rf{jt}.

\vspace{.3in}

We have noted below \rf{bmbmd} that the expectation value of the full current \rf{crdcu} over the state $|0,K>$ vanishes
since the particles and anti-particles are populated in the same manners in $|0,K>$.
In this regard, it is useful to consider, e.g., the expectation value of
\bea
j_{\m,k}(x)&\equiv & i\,[\pa_\m \f_k^*(x)\,\f_k(x)-\f_k^*(x)\,\pa_\m \f_k(x)]
\la{jmk}
\eea
\bea
\f_{k}(x)&\equiv & b_{+}^R(k) u_{k}^R(x)   +b_{-}^R(k)^{\dagger} u_{k}^{R*}
\eea
over a one-particle state such as $b_+^{R}(k)^\dagger|0,K>$:
\bea
<0,K|b_+^R(k)\,j_{\m,k}(x)\,  b_+^{R}(k)^\dagger|0,K> \la{nvc}
\eea
One gets a non-vanishing result that can easily be checked.

\subsection{Generalization to higher dimensional cases}

The discussion in the previous subsection can straightforwardly be generalized to
3D or 4D. Let us consider the 4D case to be specific.
The Schwarzschild metric is
\bea
ds^2=-\Big(1-\fr{2M}{r}\Big)dt^2+\Big(1-\fr{2M}{r}\Big)^{-1}dr^2+r^2 d\Omega^2
\eea
Setting the solution of the Klein-Gordon equation to
\bea
\fr1{r}f(r,t) Y_{lm}(\Omega)
\eea
the asymptotic form of $\f$ turns out to be (see, e.g., \cite{Birrell})
\bea
\fr1{r}Y_{lm}e^{-iwu}\quad,\quad \fr1{r}Y_{lm}e^{-iwv}
\eea
where $r$ is viewed as $r=r(r^*)$, and can be expressed in terms of the tortoise-null coordinates $(u,v)$.\footnote{
 4D extension of the observation around \rf{coneq} is not entirely straightforward due to the fact that 4D analogues of \rf{ukr} and \rf{ukl} are unknown.
For the leading $r$-behavior, the angular derivatives do not contribute;
disregarding the angular part, the 4D case reduces to the 2D case.
}
The mode expansion of $\f$ in the tortoise-null coordinates is
\bea
\f=\sum_{\w lm}\Big[b_{+,\w lm}^L u_{\w lm}^L+b_{-,\w lm}^{L\dagger} u_{\w lm}^{L*}
    + b_{+,\w lm}^R u_{\w lm}^R+b_{-,\w lm}^{R\dagger} u_{\w lm}^{R*}\Big]   \la{rfcorrect4D}
\eea
where we have defined
\bea
u_{\w lm}^R \equiv \fr1{r}Y_{lm}e^{-iwu}\quad,\quad u_{\w lm}^L \equiv  \fr1{r}Y_{lm}e^{-iwv}
\eea
Let us consider the following current for $x \in R$
\bea
j_{\m,\w lm}^+(x)\equiv i\,[\pa_\m \f_{+,\w lm}^*\,\f_{+,\w lm}-\f_{+,\w lm}^*\,\pa_\m \f_{+,\w lm}]
\eea
with
\bea
\f_{+,\w lm}\equiv   b_{+,\w lm}^R u_{\w lm}^R   \la{rfcorrect4D}
\eea
The expectation of the current goes
\bea
<0,K| j_{r^*,\w lm}^+(x) |0,K>\simeq \, \fr{2\w}{r^2}|Y_{lm}|^2
\eea
The current $j_{r,\w lm}^{+'}(x')$ in the Schwarzchild coordinates, $x'^\m=(t,r,\Omega)$, can be computed in the same manner as before:
\bea
&& <0,K|j_{r,\w lm}^{+'}(x')|0,K>
= <0,K| \fr{\pa x^\n}{\pa r}j_{\n,\w lm}^{+}(x) |0,K>\nn\\
&  \simeq & \Big(1+\fr{2M}{r}\Big) <0,K|j_{r^*,\w lm}^{+}(x)|0,K>
  \simeq  \Big(1+\fr{2M}{r}\Big) \fr{2\w}{r^2}|Y_{lm}|^2
\eea
The scattered wave should be represented by the subleading term; multiplying the surface element of dimension two, $dA_{2}\simeq r^2 d \Omega^2$, one gets
\bea
<0,K| j_{r^*,\w lm}^+(x) |0,K> dA_2\;
\simeq \sim 2\w|Y_{lm}|^2+\fr{4\w M}{r} |Y_{lm}|^2 d\Omega^2
\eea
The second term represents the flux loss for the 4D case.

\subsection{Speculation on information bleaching and escape }

In this section, we speculate on two possible modifications of the picture based on the
semiclassical description. Although the tool that we have adopted is semiclassical itself, we
try to predict what might go different in the description in which the geometry is treated as
fluctuating \cite{Park:2013iqa},\cite{Park:2013vpa}, \cite{appear}.

The results of the previous subsection indicate that there is a temporary flux loss in the outcoming wave.
The rest of the information, i.e., information on momentum of an incoming particle, would be released, and the flux loss would be compensated by subsequent thermal Hawking radiation.

The BH will undergo the thermalization process between
 the moment it absorbs the incoming wave and the moment it begins emitting Hawking
radiation. The setup that we have adopted is inadequate for analyzing what happens to the BH itself.
The precise understanding of the thermalization process
will require the exact form of the interacting theory although estimations of
certain quantities - such as the time that the BH takes to release the information - should be possible through simple models (see the discussion in \cite{Hayden:2007cs} for example).

{We believe that there are two questions whose answers might hold the key to the information paradox.}\footnote{Our main proposal - which concerns scattering of elementary states - has already been stated in previous sections, and does not sensitively depend on the validity of the speculations below.}
{The first question concerns the state of the incoming particles before they enter the BH whereas the second question concerns the state of the BH once the incoming particles enter the BH.}
{The first question is with regards to potential existence of information bleaching.
Conventionally absence of bleaching of information is assumed at the event horizon\cite{Preskill:1992tc}.
In retrospect of our result, however, we note the potential presence of certain information bleaching mechanism and its relevance for the information paradox. This is because ``what comes out" of the scattering display the information on the incoming state as stated in one of the previous subsections. (Recall that the particles are playing the role of a chair; only thermal radiation comes out in the semiclassical description of the chair disintegration process, and our result suggests that this may not be the case in the quantum description of the geometry.) Perhaps what enters the BH is in a state deprived of certain information of the incoming state. Once one considers a QFT interaction, the possibility becomes more serious since the incoming particle would make jets through the QFT interaction before it enters the BH, and they should reveal information about the incoming particles.

If we were quantitatively dealing with an interacting QFT (that results, e.g., from the ADM reduction procedure),
analyzing a {\em bound- or composite- state} scattering around a BH would be a good direction to pursue next along the line of sec 3.1. For a qualitative (and speculative) discussion, one can consider a much more complex system such as a chair thrown into a BH to bring various issues home. Eventually the BH would grow in mass after the chair entered.
In the full quantum theory of a BH, the flux/matter absorption would cause the BH to make a transition to an excited state with higher energy. Then it would go through a certain intermediate process, before it becomes a larger BH. One of the two key questions mentioned above should be whether or not the original BH remains ``black" {\em throughout} the thermalization process. The chair would be completely disintegrated at some point along the trajectory
inside the BH, and therefore would escape the ``meta BH" if the original BH does not remain entirely black during the process.\footnote{{\em As far as we are aware, there has been no proof so far that that the original black hole
remains entirely black while a fallen chair disintegrates inside it.
}} Quantum effects on the geometry must make the boundary of the BH (i.e., event horizon) smear and fluctuate. Quantum mechanical tunneling effects \cite{Parikh:1999mf}\cite{Braunstein:2011gz} should be important for the escape.\footnote{One can consider a relatively simple bound state instead of a chair, and similar effects are expected. Discussion of such a system will be more directly motivated by the main analysis of the previous sections. We have chosen a macroscopic system to make the quantum effects more drastic.}} {It may also be related to the concept of the trapping horizon \cite{Hayward:2005gi}\cite{Nielsen:2010gm}.}

\section{Conclusion}

In this work, we have employed a 2D free complex scalar theory in the background of
the 2D analogue of a Schwarzschild black hole. We have integrated the quantum mechanical
formulation of potential scattering  into the quantum field theoretic setup.
The connection between the two formulations has been leveraged by the current.
We have noted the need for introducing a distance-dependent cross section whose limiting form
 should be an analogue of the asymptotic cross section of a flat-space theory.

Our results show that flux loss occurs, and the cross section is a decreasing function
of the ``radial distance" $r$. We have interpreted this as a mass growth mechanism of the BH.
The scattering process should be unitary as a whole, and the scattered wave exhibits the
characteristics of the incoming wave. The flux loss should be temporary, and would be
compensated by subsequent Hawking radiation.
We have also speculated
on an information-escape mechanism.

The approach that we take describes events that take place outside the BH, and cannot
adequately handle what happens to the BH itself.
In the full quantum theory of a BH, the flux loss would cause the BH to make a transition into an excited state with a higher energy.
A full interacting quantum theory of gravity would also be required to understand, e,.g, the microscopic process of BH thermalization. A bound-state or a composite operator would be a good object to analyze in the extension of the current setup.
Many new effects would need to be considered once QFT interactions/non-perturbative effects are taken into account.
The presence of interactions would lead to production of particles as the original incoming particle accelerates towards the event horizon and singularity. The jets so produced may reveal information about the incoming particles.

\vspace{.2in}
There are several future directions. A direction relatively easy to pursue would be adding a QFT interaction such as $\f^4$ to the 2D model that we have considered in this work.
With an interaction present, loop-effects will be important. It is expected that jets will be produced in the vicinity of the event horizon, and information on the incoming would be revealed in much more drastic way.
Extension to a 4D case would be far more complicated once a QFT interaction is considered.
Unlike the 2D case, the solution to the quadratic partial differential equation is not known, and this would be
one of the main sources for the complication. The possibility of an information-bleaching mechanism should also be interesting to study. We will report on some of these issues in the near future.

\vspace{1in}
\ni {\bf Acknowledgements}

\ni I thank the members of CQUeST, Sogang University. Part of this work was carried out during my visit
to CQUeST.


\end{document}